\documentstyle[epsfig]{aipproc}

\begin{document}
\title{X-ray Spectral Properties of $\gamma$-ray Bursts}

\author{Tod E. Strohmayer$^1$, Edward E. Fenimore$^2$, 
Toshio Murakami$^3$ and Atsumasa Yoshida$^4$}
\address{$^1$LHEA, NASA/GSFC, Greenbelt, 
MD 20771\thanks{USRA research scientist}\\
$^2$Los Alamos National Laboratory, Los Alamos, NM, 87545\\
$^3$ISAS, 1-1 Yoshinodai 3, Sagamihara, Kanagawa, 229 Japan\\$^4$RIKEN, 2-1,
Hirowawa, Wako, Saitama, 351-01, Japan}
\maketitle
\begin{abstract}
We summarize the spectral characteristics of a sample of 22 bright
$\gamma$-ray bursts detected with the $\gamma$-ray burst sensors 
aboard the satellite {\it Ginga}. This instrument employed a
proportional and scintillation counter to provide sensitivity to photons in 
the 2 - 400 keV range, providing a unique opportunity to 
characterize the largely unexplored X-ray properties of $\gamma$-ray bursts.
The photon spectra of the {\it Ginga} bursts are well described by
a low energy slope, a bend energy, and a high energy slope.
In the energy range where they can be compared, this result 
is consistent with burst spectral analyses obtained from the BATSE 
experiment aboard the {\it Compton Observatory}. However, below 20 keV we find
evidence for a positive spectral number index in approximately
40\% of our burst sample, with some evidence for a strong rolloff
at lower energies in a few events.
We find that the distribution of spectral bend energies extends below 10 keV.
The observed ratio of energy emitted in the X-rays 
relative to the $\gamma$-rays
can be much larger than a few percent and, in fact, is sometimes larger than
unity. The average for our sample is 24\%. 
\end{abstract}

\section*{Introduction}

Twenty-five years after their discovery, $\gamma$-ray bursts (GRB) continue to
defy explanation. Analysis of burst energy spectra remains one of the principal
methods for determining the physical processes responsible for these events.
Results from the Burst and Transient Source
Experiment (BATSE) on the {\it Compton Gamma Ray Observatory} (CGRO) have
demonstrated the diversity of GRB spectral continua in the $\approx 30 - 3000$
keV range \cite{B93}, however, few instruments to
date have probed the X-ray regime of GRB between 2 and 20 keV.
Based on the first detections of X-rays in the 1-8 
keV range from GRB \cite{Wheat}, the GRB detector (GBD) 
flown aboard the {\it Ginga} satellite
was specifically designed to investigate burst spectra in the X-ray regime
\cite{Mur89}. {\it Ginga} was launched in February of 1987, and
the GBD was operational from March, 1987 until the reentry of the spacecraft in
October, 1991. Several important results have emerged from the study of burst
spectra recorded with the GBD. For example, {\it Ginga} has observed
X-ray tails in a number of bursts, as well as X-ray preactivity in one event
\cite{Yos89,Mur92}. More recently,
Beppo-Sax has observed several bursts in
both X-rays and $\gamma$-rays and discovered soft
X-ray afterglows \cite{Piro,Costa}. This latter discovery has opened the
way for the long sought GRB counterparts, 
including one with a measured redshift
\cite{Metz}.

Before BATSE, the analysis of GRB spectra suggested that their
continua could be fit by a range of models, with power law,
optically thin thermal bremsstrahlung and 
thermal synchrotron formulae providing
acceptable fits to many spectra \cite{Mazets,Hurley}. It is not at all clear,
however, that the corresponding physical processes are responsible for the
observed spectra.

\section*{Instrument Summary and Analysis}

The GBD consisted of a proportional counter (PC) 
covering the 2-25 keV range and
a scintillation counter (SC) recording photons 
between 15-400 keV. Each detector
had an $\approx$ 60 cm$^2$ effective area.  In burst mode the GBD recorded
spectral data at 0.5 s intervals for 16 s prior to and 48 s after the trigger
(MRO data). In the event that MRO data was not available for a burst, we used
the spectral data from the ``real time'' telemetry modes. For these bursts,
spectral data were available with either 2, 16, or 64 second accumulations. For
most bursts a linear fit to the MRO data in each energy channel provided a
reasonable fit to the background. Events with large variations
in the background were rejected from our analysis. For several events, the
background was estimated from real-time data. The background subtracted
spectra were then fitted using a standard $\chi^2$ minimization technique. 
We have adopted the spectral
model employed by \cite{B93} because of its relative simplicity and ability to
accurately characterize a wide range of spectral continua, in addition this 
choice facilitates direct comparison of our 
results with BATSE bursts. This model has the form
\begin{equation}
N(E) = A \left ( \frac{E}{100 \;{\rm keV}} \right )^{\alpha} 
\exp (-E/E_0) \;\;\; 
, \;\;\; (\alpha -\beta)E_0 \ge E \nonumber
\end{equation}
\begin{equation}
N(E) = A\left [ \frac{(\alpha -\beta)E_0}{100 \;{\rm keV}} \right ] ^{\alpha 
-\beta} \exp (\beta -\alpha) \left ( \frac{E}{100 \;{\rm keV}} \right )^{\beta}
\;\;\; , \;\;\; (\alpha -\beta)E_0 \le E ,
\end{equation}
where $A$ is an overall scale factor, $\alpha$ is the 
low energy slope,
$\beta$ is the high energy slope and $E_0$ is the exponential cutoff 
or bend energy. From the $\approx 120$ GRB identified by 
the GBD \cite{Og}, we selected for analysis 22 bright events for which
good spectral data were available. 
\begin{figure}[b!] 
\centerline{\epsfig{file=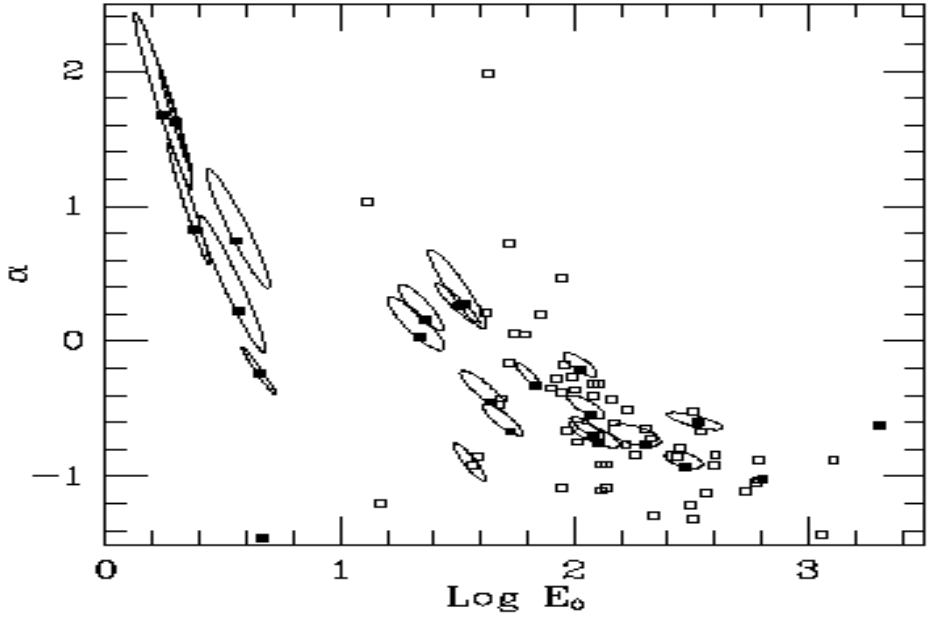,height=3.3in,width=5.0in}}
\vspace{10pt}
\caption{The 68\% confidence regions
of $\alpha$ and $E_0$ for 20 {\it Ginga} GRB. 
The confidence regions include both the effects of the 
unknown incidence angle and counting statistics. 
The solid squares denote the {\it Ginga} best-fit values 
and the open squares give 
the BATSE results from [1].}
\label{fig1}
\end{figure}
Sky positions
for four of the events in our sample are known because of simultaneous
detections with either BATSE or WATCH. For the remaining events, the incidence
angle $\theta$ of the photons into the 
detectors is uncertain ($0^{\circ}$ equals normal incidence). Since the 
detector response is a function of this angle, the inferred 
source spectrum and thus peak intensity of these events is also somewhat
uncertain. We selected $37^{\circ}$ for the incidence angle when the angle was
unknown. This is a typical angle considering that the mechanical support
for the window on the PC acts as a collimator limiting the field of view to
an opening angle of $\sim \pm 60$ degrees.  We used
Monte Carlo simulations to evaluate the impact of the unknown incidence
angle. For each simulation, we selected a random incidence angle between
0 and 60 degrees.  We calculated a response matrix and used it with the burst's
best fit parameters to generate simulated spectra.  Background
was added and Poisson statistics applied.  We then analyzed the simulated
data the same way we analyzed the burst data: an estimated background
was subtracted and the best fit parameters were
found based on a response matrix corresponding to $37^{\circ}$. Based on these
simulations we conclude that the lack of knowledge of the incidence angle into
{\it Ginga} does not introduce much more uncertainty than the counting
statistics. On average, because of the uncertain incidence angle, 
the confidence
region for $E_0$ is 22\% larger and the confidence region 
for $\alpha$ is larger
by 0.06 than that produced by counting statistics alone ({\it cf.} \cite{Stroh}
for details of the simulations).

\section*{X-ray Spectral Characteristics of GINGA Bursts}

Our spectral fits are generally acceptable, with $\chi^2_r$ of order unity for
most of the bursts. In agreement with \cite{B93} we find that a range in the
model parameters is required to adequately describe GRB spectra.
Of particular interest is the behavior of the {\it Ginga} 
sample at X-ray energies. About 40\%  of the bursts in the sample show a 
positive spectral slope below 20 keV (i.e., $\alpha > 0$), with the
suggestion of rolloff toward lower energies in a few of the bursts ($\alpha$ as
large as $\sim +1.5$). Unfortunately, the lack of data below 1 keV, and the
often weak signal below 5-10 keV precludes us
from establishing the physical process (photoelectric absorption,
self-absorption) that may be involved in specific bursts. 
The remainder of the burst spectra continue to increase below 10 keV.
Observations of the low-energy asymptote can place constraints
on several GRB models, most notably the synchrotron shock model which
predicts that $\alpha$ should be between -3/2 and -1/2 \cite{Katz}.
BATSE data have been used to argue that some GRB
violate these limits during some {\it time-resolved} samples \cite{Crider}. We
find violations of these limits in the {\it time-integrated} events.

In Figure \ref{fig1}, we show the 68\%
confidence regions for 20 of the bursts in our sample. These confidence 
regions include our estimate of the effects of the uncertain incidence angle.
For two bursts we only show the best fit parameters as solid squares because a
simple power law could nearly fit the entire spectra. In Figure \ref{fig1} we
also show the new {\it Ginga} results with 
those of \cite{B93}. The open squares
are the BATSE results and the solid squares are the {\it Ginga}
results. Many of the {\it Ginga} points lie within the range found by BATSE.
However, the lowest $E_0$ found by BATSE was 14 keV (set, of course, by the
lowest energy observed by BATSE).  Ginga extends $E_0$ values down to 2
keV.  BATSE had a small fraction (15\%) of events with $\alpha > 0$
whereas {\it Ginga} has 40\% of events with $\alpha > 0$.
In general this is because there is a correlation between $\alpha$ and
$E_0$ such that the lower energy range of {\it Ginga} samples a parameter
space with more events with $\alpha > 0$. For the 76 points, the Pearson's
$r$ coefficient is -0.62.
The formal significance is about 4$\sigma$ although that
ignores the complicated error bars that
are caused by the fact that the observations tend to agree with a range
of $\alpha-E_0$. However, the existence of the correlation 
seems reasonable: there are virtually no
events seen by BATSE at large $\alpha$, large $E_0$ and few low $\alpha$,
low $E_0$ events seen by {\it Ginga}.

In our sample of bursts, we find that $\alpha$ can be both positive and
negative. Negative $\alpha$'s are often seen in
time-integrated BATSE spectra.  Positive $\alpha$'s where the spectrum
rolls over at low energies are usually only seen in {\it time
resolved} BATSE spectra \cite{Crider}.
 
The {\it Ginga} trigger range (50 to 400 keV) was virtually the same
as BATSE's. Thus, we do not think we are sampling a different population of
bursts, yet we get a different range of fit parameters.
The lack of events with $E_0$'s between 6 and 20 keV cannot be used to support
two populations because we do not have enough events.
One possible explanation might be that GRBs have two break energies, one
often in the 50 to 500 keV range and the other near 5 keV.  Both BATSE and
{\it Ginga} fit with only a single break energy so BATSE tends to find breaks
near the center of its energy range and we tend to find
breaks in our energy range.
Without good high energy observations of bursts with low $E_0$, it is
difficult to know whether they also have a high energy bend.

\cite{Preece} utilized a BATSE low energy discriminator channel and detected
emission in excess of what would be expected from a fit at higher energy.
They report excesses in 15\% of the investigated BATSE
bursts. One of our bursts, GB880205, shows a clear strong excess at low energy
and two other {\it Ginga} bursts, GB880830 and GB910418,
probably also show an excess.
For GR910418, \cite{Preece} also reported 
an excess. From \cite{Preece} we would
expect about 3 of our bursts to show an excess so we are consistent with the
BATSE result.

\end{document}